\documentclass[twocolumn,prc,aps,showpacs,amsmath,amssymb,nofootinbib]{revtex4}
\usepackage{graphicx}

 \def\Pom{{ I\!\!P}}
 \def\Reg{{ I\!\!R}}
 \newcommand\la{\langle}
 \newcommand\ra{\rangle}
 \newcommand\beq{\begin{equation}}
 
 \newcommand\eeq{\end{equation}}
 \newcommand\beqn{\begin{eqnarray}}
 \newcommand\eeqn{\end{eqnarray}}
\def\mb{\,\mbox{mb}}
\def\fm{\,\mbox{fm}}
\def\GeV{\,\mbox{GeV}}

\def\lsim{\mathrel{\rlap{\lower4pt\hbox{\hskip1pt$\sim$}}
    \raise1pt\hbox{$<$}}}         
\def\gsim{\mathrel{\rlap{\lower4pt\hbox{\hskip1pt$\sim$}}
    \raise1pt\hbox{$>$}}}         

\def\Im{\,\mbox{Im}\,}
\def\mb{\,\mbox{mb}}
\def\fm{\,\mbox{fm}}
\def\GeV{\,\mbox{GeV}}

\def\s0{\sigma_0(s)}
\def\sq{\sigma_{\bar qq}}

\def\Im{\,\mbox{Im}\,}
\def\mb{\,\mbox{mb}}
\def\fm{\,\mbox{fm}}
\def\GeV{\,\mbox{GeV}}

\begin{document}

\title{\bf Evidences for two scales in hadrons}

\author{B.Z.~Kopeliovich$^{a,b}$}
\author{I.K.~Potashnikova$^{a}$}
\author{B.~Povh$^{c}$}
\author{Ivan~Schmidt$^a$}

\affiliation{$^a$Departamento de F\'\i sica
y Centro de Estudios
Subat\'omicos,\\ Universidad T\'ecnica
Federico Santa Mar\'\i a, Casilla 110-V, Valpara\'\i so, Chile\\
$^b$Joint Institute for Nuclear Research, Dubna, Russia\\
$^c$Max-Planck-Institut f\"ur Kernphysik, Postfach 103980,
69029 Heidelberg, Germany}

\date{\today}

 \begin{abstract}

Some unusual features observed in hadronic collisions at high
energies can be understood assuming that gluons in hadrons are
located within small spots occupying only about 10\% of the hadron's
area. Such a conjecture about the
presence of two scales in hadrons helps to explain:\\
why diffractive gluon radiation so much suppressed;\\
why the triple-Pomeron coupling shows no t-dependence;\\
why total hadronic cross sections rise with energy so slowly;\\
why diffraction cone shrinks so slowly, and why 
$\alpha^\prime_\Pom\ll\alpha^\prime_\Reg$;\\
why the transition from hard to soft regimes in the structure
functions occurs at rather large $Q^2$;\\
why the observed Cronin effect at collider energies is so weak;\\
why hard reactions sensitive to primordial parton motion (direct photon,
Drell-Yan dileptons, heavy flavors, back-to-back di-hadrons, seagull
effect, etc.) demand such a large transverse momenta of the projectile
partons, which is not explained by NLO calculations;\\
why the onset of nuclear shadowing for gluons is so much delayed
compared to quarks, and why shadowing is so weak.

\end{abstract}

\pacs{12.38.Aw, 12.38.Qk, 12.40.Nn, 12.38.Bx}

\maketitle

\section{Introduction}

The first manifestations of the gluon contribution to the total
hadronic cross sections go back to the early 70s when energy rising
cross sections were observed in ISR experiments. Indeed, treating a
hadron as a bound state in nonrelativistic quantum mechanics, its
transverse size is invariant relative to longitudinal boosts, and
therefore one may expect a constant cross section at high energies
(Pomeranchuk limit). Rising cross sections require new degrees of
freedom, which was attributed to the emission of gluons.

The radiation of gluons is quite different from photon radiation.
Gluons can radiate other gluons, and therefore with the same
probabilities at any rapidity, which makes it easier for gluons to
cover the large rapidity interval between the colliding hadrons. On
the contrary, photon radiation vanishes exponentially at large
rapidity intervals from the source. Due to integration over phase
space the radiation of each gluon brings a power of $\ln(s)$ to the
cross section (BFKL regime \cite{bfkl}). Another clear difference
between QCD and QED is that the non-Abelian nature of the
interactions results in elastic hadronic amplitudes which are nearly
imaginary (nearly real in QED).

One may wonder however why the observed rise of the total hadronic
cross sections is so slow, approximately $s^{0.1}$? The total cross
section is dominated by soft interactions, so the QCD coupling is
large, and a steep energy dependence is in principle expected. Some
dynamics must therefore suppress gluon radiation. This observation
might have been a first hint to the idea of the existence of a
semihard scale in hadrons, which controls and suppresses gluon
radiation.

Another hint to this phenomenon came by in about the same period of
time, from data on diffraction. It was discovered that diffractive
gluon radiation is quite suppressed compared to simple expectations
for this soft reaction. This is the well known problem of the
smallness of the triple-Pomeron coupling, which is a direct
indication of the presence of a semihard scale.

Thus, data suggest an enhanced Fermi motion of gluons in light
hadrons compared to the inverse hadronic radius. Correspondingly,
gluons in the hadron should be located within spots of a small size.
This picture was suggested in \cite{kst2}, and the mean radius of
the spots was fitted to diffractive data at $r_0=0.3\fm$. Therefore
the area of a gluonic spot is an order of magnitude smaller than the
area of the hadron.

This observation is in accord with nonperturbative QCD models. For
instance, a small gluon correlation radius $0.35\fm$ is a result of
lattice calculations \cite{lattice}, and it is also predicted by the
instanton model \cite{shuryak}, related in this case to the
instanton size $0.3\fm$. Phenomenological applications of this
effect were discussed in \cite{shuryak-zahed}. The smallness of the
mean radius of gluon distribution in the proton was also confirmed
by an analysis of hadronic matrix elements of the gluonic
contribution to the energy momentum tensor, using QCD sum rules. It
gave a value of 0.3 fm for the radius of the corresponding gluonic
form factor \cite{braun}.

In this paper we present several experimental facts, which so far
have not had a clear interpretation, but which can be explained
assuming the location of gluons within small spots. In
Sect.~\ref{diffraction} we demonstrate that diffractive gluon
bremsstrahlung is strongly suppressed, and that this fact is an
important evidence for the smallness of gluonic spots. Such a
conjecture also explains an unusual feature of the triple Pomeron
coupling, which is the absence of a form factor.

In Sect.~\ref{sig-tot} we relate the smallness of the energy slope,
$\epsilon=d\ln(\sigma_{tot})/d\ln(s)=0.1$ to the gluonic semihard scale in the
proton.

In Sect.~\ref{slope} we consider the longstanding problem of the smallness of
the Pomeron trajectory slope, $\alpha^\prime_\Pom=0.25\GeV^{-2}$, compared to
Regge trajectories, $\alpha^\prime_\Reg=0.9\GeV^{-2}$. Within the same model
of gluonic spots we predict an even smaller value
$\alpha^\prime_\Pom=0.1\GeV^{-2}$, which indeed was observed in
electroproduction of vector mesons. Although the effective slope of the
Pomeron trajectory observed in $pp$ collisions,
$\alpha^\prime_\Pom=0.25\GeV^{-2}$, is larger than expected, this is well
explained by saturation of unitarity. Within the same model we well
reproduced the large value of the slope of Regge trajectories,
$\alpha^\prime_\Reg\approx 1\GeV^{-2}$.

In Sect.~\ref{hard-soft} we study the manifestations of the two
hadronic scales in the transition from hard to soft regimes in DIS.
We conclude that the observed slowing down towards the soft limit
and final drop of the logarithmic $Q^2$ derivative of the structure
function is related not to the phenomenon of saturation expected at
small $x$, but to the reduction of $Q^2$ correlated with $x$ in
data.

Naturally, gluons located within small spots have an enhanced Fermi
motion. In Sect.~\ref{primordial} we list experimental evidences for
an increased primordial transverse momenta of gluons in projectile
hadrons. One of the consequences is a very weak Cronin effect,
predicted in \cite{knst} and successfully verified at RHIC.

Localization of gluons within small spots leads also to dramatic
changes in nuclear effects of coherence, considered in
Sect.~\ref{nuclei}, such as gluon shadowing and color glass
condensate, which are expected to be much weaker than usually
believed. This is a result of the rather small overlap of the
gluonic spots in the transverse plane. Increased transverse momentum
of gluons also makes them effectively heavier. This results in a
reduction by an order of magnitude of the coherence time related to
gluon shadowing, and delays the onset of shadowing of gluons
compared to quarks.

Some of the listed results have been already published in conference
proceedings \cite{kp,kps}.

\section{Gluon radiation suppression}\label{diffraction}

If gluons in hadrons are located in small spots of radius $r_0$,
they have enhanced transverse momenta $q_T\sim 1/r_0$. For soft
interactions it is difficult to resolve such gluons and shake them
off, i.e. the gluon bremsstrahlung cross section should be
suppressed compared to perturbative estimates.  However, in the case
of an inelastic collision followed by multiparticle production the
events with or without gluon radiation look alike. Indeed, a
color-exchange interaction results in two flying away colored
clusters of remnants of the colliding hadrons, which can produce
particles even if no gluons are radiated, just by means of string
breaking. The produced particles build a plateau in rapidity similar
to gluon bremsstrahlung, and it is therefore difficult to find any
certain signature for the radiated gluons.

\subsection{Small triple-Pomeron coupling}

On the other hand, diffraction offers an exceptional possibility to
identify gluon radiation. A high-energy hadron can dissociate
diffractively either via excitation of the valence quark skeleton,
or by the radiation of gluons, and these two mechanisms are
characterized by different dependence on the effective mass ($M_X$)
of the excitation \cite{kklp},
 \beqn
\frac{d\sigma(hp\to X p)}{dM_X^2} =
\left\{
\begin{array}{lc}%
\frac{1}{M_X^3} & {\rm quark\ skeleton}\\
\frac{1}{M_X^2} & {\rm gluon\ bremsstrahlung}
\end{array}
\right.
\label{100}
 \eeqn
 The $M_X$-dependence at large $M_X$ correlates with the spin of the
slowest particle produced in the excitation. Only a vector particle,
i.e. a gluon, can provide a $1/M_X^2$ dependence.  Thus, one can
single out the large mass tail from the $M_X$-distribution which
gives the cross section of diffractive gluon radiation.

\begin{figure}[htb]
 \includegraphics[width=7cm]{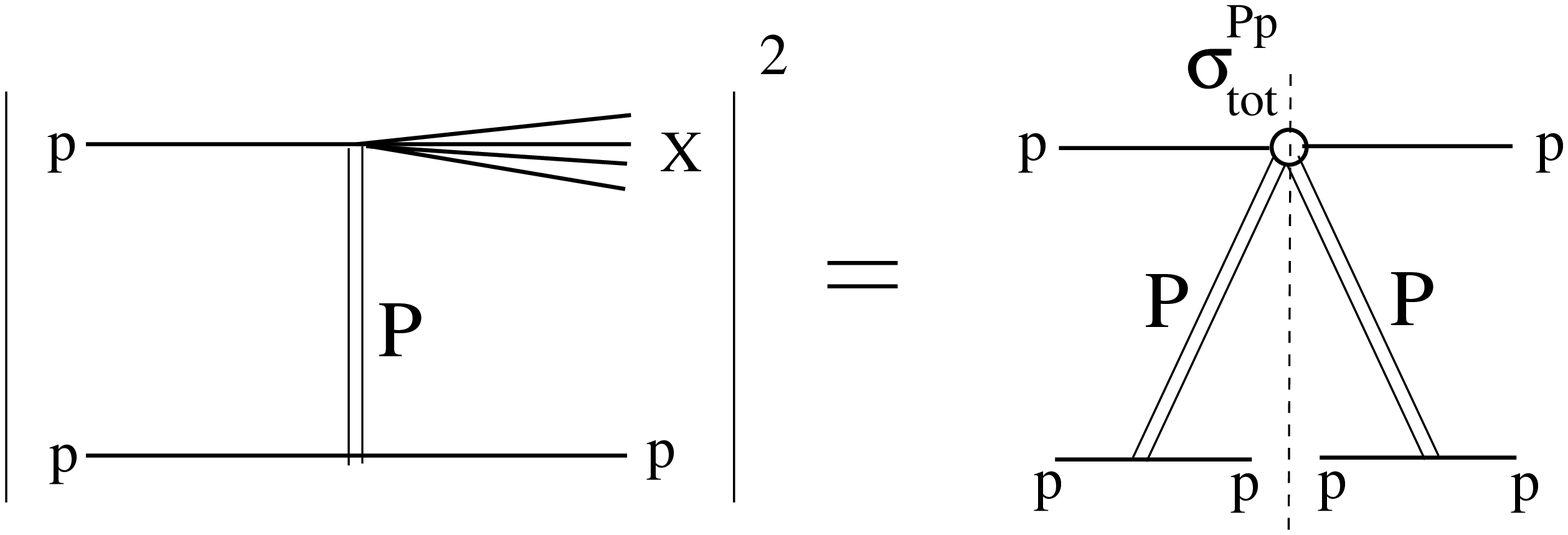}
\caption{\label{3R}
The cross section of diffractive excitation
of a proton expressed in terms of the total Pomeron-proton cross section.}
 \end{figure}
\begin{figure}[htb]
 \includegraphics[width=8cm]{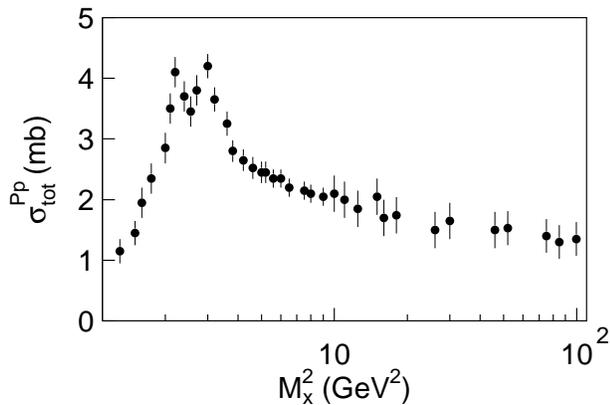}
\caption{\label{pom-p} Pomeron-proton total cross sections
$\sigma^{\Pom p}_{tot}$ as function of invariant mass squared,
$s'=M_X^2$, extracted from data on $pp\to pX$ \cite{kaidalov}.}
 \end{figure}

There is a simple hint showing that the diffractive excitation cross
section is unexpectedly small. In fact, it has been known since the
70s that the triple-Pomeron coupling is quite small. To appreciate
this statement one can express diffraction in terms of the
Pomeron-proton total cross section, as is demonstrated pictorially
in Fig.~\ref{3R}.
 Treating the Pomeron as a gluonic dipole, one may expect a cross
section which is a Casimir factor $9/4$ larger than that for
quarkonium. Comparing with the pion-proton cross section one arrives
at about $50\mb$. However, experimental results for $\sigma^{\Pom
p}_{tot}(s'=M_X^2)$, depicted in Fig.~\ref{pom-p}, give a quite
smaller value, less than $2\mb$ at large $M_X^2$, where the triple
Pomeron term is expected to dominate.
 At smaller invariant masses $M_X$ the triple Regge term $\Pom\Pom\Reg$
is responsible for the contribution that decreases with decreasing
$M_X$.

\begin{figure}[htb]
 \includegraphics[width=7cm]{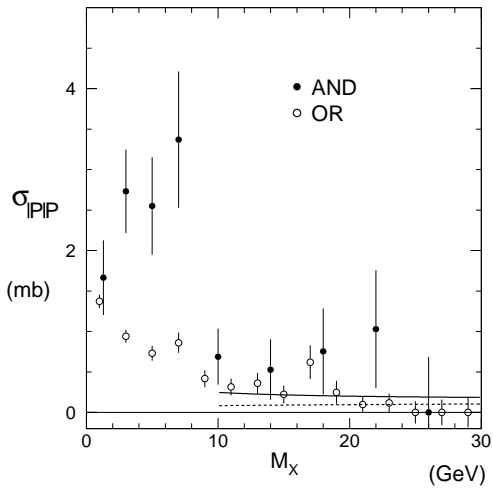}
\caption{\label{pom-pom} The total Pomeron-Pomeron cross section
measured in double diffractive processes \cite{schlein1}. Samples
labeled "AND" and "OR" correspond to different criteria of event
selection. The solid and dashed curves correspond to expectations
based or Regge factorization \cite{schlein1}. }
 \end{figure}

The Pomeron-Pomeron total cross section measured in double diffractive
processes is plotted in Fig.~\ref{pom-pom}.
 It turns out that $\sigma^{\Pom\Pom}_{tot}$ is even an order of
magnitude smaller than $\sigma^{\Pom p}_{tot}$.

Of course a glue-glue dipole should interact stronger than a $\bar
qq$ dipole of the same size. However, if the former has a smaller
transverse size, then its interaction is subject to color
transparency \cite{zkl} and the cross section may be very small.
Therefore a straightforward explanation for the above dramatic
disagreement would be a much smaller size of the gluonic dipole
(Pomeron) compared to the quark-antiquark dipole (pion), and as a
consequence one concludes that gluons should be located within small
spots in the proton. Moreover, making the simple conjecture that the
cross section is $\pi R_\Pom^2$, the Pomeron radius was found in
\cite{ingelman} to be $R_\Pom^2=0.01\fm^2$, which is an order of
magnitude smaller than the result of \cite{kst2} and the evaluations
done in this paper.

The above argumentation is rather speculative, since the possibility
to describe diffraction in terms of a Pomeron flux is questionable
beyond the Regge-pole model. A more consistent approach would be a
calculation of gluon radiation caused by a colorless gluonic
exchange (Pomeron). If Weizs\"acker-Williams gluons in the proton
are located within small spots which are hardly resolved by soft
interactions, bremsstrahlung will be suppressed. The mean transverse
size of these spots was fitted in \cite{kst2} to single diffraction
data with large effective masses and found to be $\la r_T\ra\sim
r_0=0.3\fm$. Such gluons have much more intensive Fermi motion than
massless perturbative ones, and they are less sensitive to an
external kick, i.e.  gluon radiation is suppressed.

Let us start by calculating the  cross section of diffractive gluon
radiation by a high-energy quark interacting with a nucleon. For
this purpose we need the light-cone wave function of the quark-gluon
Fock component of a quark, which was calculated in \cite{kst2}
within a model describing the nonperturbative interaction of gluons
via a phenomenological light-cone potential taken in an oscillatory
form. The result reads,
 \beq
\Psi_{qg}(\vec r_T)= -\frac{2i}{\pi}\,
\sqrt{\frac{\alpha_s}{3}}\
\frac{\vec r_T\cdot\vec e^{\,*}}{r_T^2}\,
{\rm exp}\left(-{r_T^2\over2r_0^2}\right)\ .
\label{302}
 \eeq
 Here $r_T$ is transverse quark-gluon separation; $\vec e$ is the gluon
polarization vector. We assume that the gluon, which is a vector particle, is
carrying a negligible fraction $\alpha\ll1$ of the quark momentum.

Of course the concrete shape of the light-cone potential is not crucial, 
important is the smallness of the mean quark-gluon separation.
Alternatively, one can rely on the perturbative light-cone distribution
amplitude for the quark-gluon fluctuation, but assuming the gluon to be
effectively heavy. The corresponding light-cone distribution amplitude 
for radiation of a transversely polarized gluon with $\alpha\ll1$ reads
\cite{kst1,kst2},
\beq
\Psi_{qg}(\vec r_T)=-{2i\over\pi}\,
\sqrt{\frac{\alpha_{s}}{3}}\,  
\frac{\vec r_T\cdot\vec e^{\,*}}{r_T}\,
m_g\,K_1(r_Tm_g)\,,
\label{302a} 
\eeq
 With the effective gluon mass $m_g\approx 0.7\GeV$ the mean quark-gluon
transverse separations with Eqs.~(\ref{302}) and (\ref{302a}) are
similar.Thus, either Eq.~(\ref{302}), or Eq.~(\ref{302a}) can be viewed as a
source of small gluonic spots in the proton.

Once the light-cone quark-gluon distribution function is known, one
can calculate the cross section of diffractive gluon radiation by a
high-energy quark interaction with a nucleon \cite{kst2},
 \beqn
&&\frac{d\sigma(qN\to qGN)}{dx_F\,dp_T^2}
\biggr|_{p_T=0} = \frac{1}{16\pi(1-x_F)}
\nonumber\\ &\times&
\int d^2r_T\,\biggl|\Psi_{qG}(\vec r_T)\biggr|^2\,
\widetilde\sigma^2(r_T,s)\ .
\label{303}
 \eeqn
 Here the cross section $\widetilde\sigma(r_T,s)$ is not just the
cross section of interaction of a $qG$ dipole. In fact, this dipole
is not even colorless. As usual, diffractive excitation is possible
due to the difference between elastic amplitudes for different Fock
states, in this case the bare quark $|q\ra_0$ and the $|qG\ra$ pair.
Since they have the same color, the difference emerges from the
color-dipole moment of the $q-G$ pair. It is shown in \cite{kst2}
(see in particular Appendix~A.2) that
$\widetilde\sigma(r_T,s)={9\over8}\,\sq(r_T,s)$, where $\sq(r_T,s)$
is the cross section for the interaction of a nucleon with a
$q\bar{q}$ dipole of transverse separation $r_T$..

We use the saturated form of this energy dependent dipole cross
section, suitable for the soft processes under consideration,
 \beq
\sigma_{\bar qq}(r_T,s)=\sigma_0(s)\,\left[
1-{\rm exp}\left(-\frac{r_T^2}
{R_0^2(s)}\right)\right]\ ,
\label{180}
 \eeq
 where $R_0(s)=0.88\,fm\,(s_0/s)^{0.14}$ and $s_0=1000\,GeV^2$
\cite{kst2}. The energy dependent factor $\sigma_0(s)$ is defined as,
 \beq
\sigma_0(s)=\sigma^{\pi p}_{tot}(s)\,
\left(1 + \frac{3\,R^2_0(s)}{8\,\la r^2_{ch}\ra_{\pi}}
\right)\ ,
\label{190}
 \eeq
 where $\la r^2_{ch}\ra_{\pi}=0.44\pm 0.01\,fm^2$ is the mean
square of the pion charge radius. The $s$-dependent dipole cross
section Eq.~(\ref{180}) is fitted \cite{kst2} to data for hadronic
cross sections, for real photoproduction and also for low-$Q^2$ HERA
data for the proton structure function.  The cross section
(\ref{180}) averaged with the pion wave function squared (see below)
automatically reproduces the pion-proton cross section.

The next step is to integrate over $p_T$ the cross section of
diffractive gluon radiation provided that the forward one,
Eq.~(\ref{303}), is known. In terms of Regge phenomenology,
diffractive radiation corresponds to the triple-Pomeron term in the
cross section of single diffraction. Data agree with a Gaussian
$p_T$-dependence of the triple-Pomeron term with the slope
\cite{kklp},
 \beq
B^{pp}_{3\Pom}(x_F) = B_{3\Pom}^0+2\,\alpha_{\Pom}^\prime\,
\ln\left(\frac{1}{1-x_F}\right)\ ,
\label{304}
 \eeq
 where $B_{3\Pom}^0=4.2\GeV^{-2}$, and $\alpha_{\Pom}^\prime =
0.25\GeV^{-2}$.

Now we are in a position to evaluate the effective triple-Pomeron
part of the single diffraction cross section for $pp$ collisions,
employing the wave function Eq.~(\ref{302}) and the saturated shape
for the cross section Eq.~(\ref{180}),
 \beqn
&& \left[\frac{d\sigma(pp\to pX)}{dx_F\,dp_T^2}\right]_{3\Pom}
= \frac{81\,\alpha_s\,\sigma_0^2(s)}{(16\pi)^2(1-x_F)}
\nonumber\\ &\times&
\ln\left[1+\frac{\epsilon^2(s)}{1+2\epsilon(s)}\right]\,
\exp\left[-p_T^2\,B^{pp}_{3\Pom}(x_F)\right]\ ,
\label{305}
 \eeqn
 where $\epsilon(s)=r^2_0/R^2_0(s)$. At high energies $\epsilon(s)$ is
rather small, and then the single diffractive cross section
Eq.~(\ref{305}) is proportional to $r_0^4$. This is why this process
is quite sensitive to the value of $r_0$, and therefore provides an
efficient way to determine the size of gluonic spots in the proton,
$r_0\approx 0.3\fm$ \cite{kst2}.

Notice that the interference between the diffractive amplitudes of gluon
radiation by different quarks in the proton should not be appreciable,
since $r_0$ is small compared to the proton radius. Explicit calculations
performed in \cite{kst2} confirm this.

\subsection{Unitarity corrections}\label{unitarity}

Any large rapidity gap process is subject to unitarity or absorptive
corrections, which may be substantial. Indeed, initial/final state
interactions tend to fill the rapidity gap by producing particles,
and one may treat such corrections as a survival probability of the
rapidity gap. Such corrections are in general especially large and
may completely terminate the gap in the vicinity of the unitarity
limit, which is also called black disk regime.  Notice that elastic
scattering is an exclusion: unitarity corrections enhance it, and
the elastic cross section is maximal in the black disk limit.

Since the phenomenological dipole cross section $\sq(s)$ is fitted
to data, we assume that it includes by default all the unitarity
corrections. Thus, the amplitude of diffractive excitation of a
quark, $qN\to qGN$, does include all the absorptive corrections
contained in the phenomenological dipole cross section. However, the
presence of other projectile valence quarks, the spectators, should
not be ignored. Indeed, any inelastic interactions of the large-size
three quark dipole in the proton, will cause particle production
which will fill the rapidity gap. Thus one may expect large
absorptive corrections to the cross section of diffractive gluon
radiation.

Data for elastic $pp$ scattering show that the partial amplitude
$f^{pp}_{el}(b,s)$ hardly varies with energy at small impact
parameters $b\to 0$, while rises as a function of energy at large
$b$ \cite{amaldi,k2p,k3p}. This is usually interpreted as a
manifestation of saturation of the unitarity limit, $\Im
f^{pp}_{el}\leq 1$. Indeed, this condition imposes a tight
restriction at small $b$, where $\Im f^{pp}_{el}\approx 1$, leaving
almost no room for further rise. Correspondingly, the amplitude of
any off-diagonal process including single triple-Pomeron diffraction
acquires a suppression factor
 \beq
f^{pp}_{sd}(b,s)\Rightarrow
f^{pp}_{sd}(b,s)\,\left[1-\Im f^{pp}_{el}(b,s)\right]\ ,
\label{306}
 \eeq
 due to unitarity or absorptive corrections. This factor, which expresses the
survival probability of LRG, is known to decrease with energy
\cite{glm}. The interplay of the rising and falling energy
dependence of the two factors in (\ref{306}) may explain the
observed flat behavior of the single diffractive cross section
\cite{dino,peter}.

 Since $\Im f^{pp}_{el}(b,s)$ is known directly from data, it would be
straightforward to just fit the data with any proper
parametrization, and use the result in Eq.~(\ref{306}).
Alternatively, one can use any model which provides a good
description for $\Im f^{pp}_{el}(b,s)$. It is demonstrated in
\cite{k3p} that even the simple model that treats the Pomeron as a
Regge pole with no unitarity corrections, describes reasonably well
not only the total hadronic cross sections, but even the data for
$f^{pp}_{el}(b,s)$. In this model,
 \beq
\Im f^{pp}_{el}(b,s)=
\frac{\sigma^{pp}_{tot}(s)}
{4\pi B^{pp}_{el}(s)}\
\exp\left[-\frac{b^2}{2
B^{pp}_{el}(s)}\right]\ .
\label{307}
 \eeq
 Here and for further applications we use the parametrization from
\cite{pdg}, $\sigma^{pp}_{tot}(s) = 18.76\mb\times(s/M_0^2)^{0.093}
+ \sigma^{pp}_R(s)$, where $M_0=1\GeV$. The Reggeon part of the
cross section $\sigma^{pp}_R(s)$ is small at high energies and can
be found in \cite{pdg}. The elastic slope is
 \beq
B^{pp}_{el}(s) = B^0_{el} +
2\,\alpha^\prime_{\Pom}\,\ln(s/M_0^2)\,,
\label{307a}
 \eeq
 with $B^{0}_{el}=8.9\GeV^{-2}$ and $\alpha^\prime_{\Pom}=0.25\GeV^{-2}$.
Note that the elastic amplitude at small impact parameters, i.e. the
pre-exponential factor in (\ref{307}), hardly changes with energy,
imitating saturation of unitarity. This fact was known back in the
70s as a geometrical scaling. It is demonstrated in \cite{k3p} (see
Fig.~9) that not only at $b=0$, but in the whole range of impact
parameters the model Eq.~(\ref{307}) describes reasonably well the
energy dependence of the partial amplitude $f^{pp}_{el}(b,s)$.

Using the Fourier transformed Eq.~(\ref{307}) we arrive at the
following cross section for single diffraction integrated over
momentum transfer,
 \beqn
&&\left[\frac{d\sigma(pp\to pX)}{dx_F}\right]_{3\Pom}
= \frac{81\,\alpha_s\,\sigma_0^2(s)}
{(16\pi)^2(1-x_F)\,B^{pp}_{3\Pom}(x_F)}\
\nonumber\\ &\times&
\ln\left[1+\frac{\epsilon^2(s)}{1+2\epsilon(s)}\right]\,
\left\{1-{1\over{\pi}}\,\frac{\sigma^{pp}_{tot}(s)}
{B^{pp}_{3\Pom}(x_F)+2B^{pp}_{el}(s)}\right.
\nonumber\\ &+& \left.
\frac{1}{(4\pi)^2}\,
\frac{\left[\sigma^{pp}_{tot}(s)\right]^2}
{B^{pp}_{el}(s)
\left[B^{pp}_{3\Pom}(x_F)+B^{pp}_{el}(s)\right]}\right\}\ .
\label{308}
 \eeqn

Comparing this result with experimental data one can find the mean gluonic
radius $r_0$. We use the triple-Regge fit \cite{kklp} which gives for the
triple-Pomeron part of the cross section,
 \beq
\left[\frac{d\sigma(pp\to pX)}{dx_F}\right]_{3\Pom}=
\frac{G_{3\Pom}(0)}{(1-x_F)B^{pp}_{3\Pom}(xF)}\,,
\label{310}
 \eeq
 where $G_{3\Pom}(0)=3.2\mb/\GeV^2$ is the effective triple Pomeron
coupling at $t=0$, including also three Pomeron-proton vertices.

 We fix the QCD coupling at the scale corresponding to the mean transverse
momentum of gluons, $\alpha_s(1/r^2_0)\approx 0.4$ \cite{k3p}. Then,
comparing Eqs.~(\ref{308}) and (\ref{310}) we arrive at
$r_0\approx0.3\fm$.

\subsection{Why the triple-Pomeron vertex has no structure}

The elastic slope Eq.~(\ref{307a}), which is a half of the $pp$
interaction radius squared, contains two terms. The second term,
which depends on energy, has its origin in the Gribov's diffusion of
partons in the transverse plane. The first constant term,
$B^0_{el}$, comes from the the Pomeron-proton form factors of the
colliding protons.

Analogously, the first term in the triple-Pomeron slope,
Eq.~(\ref{304}), should consist of contributions coming from the
form factors of both the proton and the triple-Pomeron vertex.
Surprisingly, it turns out that $B^0_{3\Pom}\approx
{1\over2}B^0_{el}$ which is just the contribution of the target
proton. Nothing is left for the triple-Pomeron vertex! This has been
another known puzzle since the Regge era \cite{kklp}.

To evaluate how much a dipole of transverse separation $r_0$
contributes to the slope compared to the proton contribution, we use
the mean transverse diameter of the proton squared, which equals
${8\over3}\,\la r^2_{ch}\ra$, where the mean proton charge radius
squared $\la r^2_{ch}\ra=0.8\fm^2$. Then the relative correction to
the slope related to the dipole size is,
 \beq
\delta=\frac{3r_0^2}{8\la r^2_{ch}\ra}=4\%
\label{330}
 \eeq
 This very small result explains why the triple-Pomeron vertex looks
structureless in data. This is another manifestation of the presence
of small gluonic spots in the proton.

\section{Total hadronic cross sections: why the energy dependence is so
weak}\label{sig-tot}

It has been known since 1973 that hadronic cross sections rise with energy
approximately as $s^\epsilon$, where the exponent is quite small,
$\epsilon\approx 0.1$. What is the origin of this small number? We do not
expect any small parameters in the soft regime of strong interactions.

The rise of the cross sections is related to gluon bremsstrahlung.
Indeed, without gluon radiation the geometrical cross section of two
hadrons would be constant, since their transverse size is Lorentz
invariant, i.e. is energy independent. In fact, the radiation of
each gluon has a phase space proportional to $\ln(s)$, and
multigluon radiation leads to powers of $\ln(s)$ in the cross
section. On the other hand, we have already concluded that radiation
of gluons is suppressed if they are located within small spots,
since in this case it is difficult to resolve and shake them off.
Thus, the suppression of diffractive gluon radiation and the
observed weak energy dependence of total hadronic cross sections
have the same origin.

The calculations performed in \cite{k3p} confirm this. The hadronic cross
section was found to have the following structure,
 \beq
\sigma_{tot}=\sigma_0 + \sigma_1\,
\left(\frac{s}{s_0}\right)^\Delta\ ,
\label{200}
 \eeq
 where $\sigma_0$ is the energy independent term, related to hadronic
collisions without gluon radiation. The second term in (\ref{200})
is related to the contribution of gluon bremsstrahlung to the total
cross section. Since this part is expected to be small, $\sigma_1$
should be small. Indeed, it was found in \cite{k3p} that
$\sigma_1=27/4\,C\,r_0^2$, where the factor $C\approx 2.4$ is
related to the behavior of the dipole-proton cross section
calculated in Born approximation at small separations, $\sigma(r_T)=
Cr_T^2$ at $r_T\to0$.

The energy dependence of the second term in (\ref{200}) was found to
be rather steep, with an exponent $\Delta=4\alpha_s/3\pi=0.17$,
which seems to be too large compared to the experimentally measured
$\epsilon\approx 0.1$. There is, however, no contradiction due to
presence of the energy independent first term in (\ref{200}).
Approximating the cross section (\ref{200}) by the simple power
dependence on energy, then the effective exponent reads,
 \beq
\epsilon=\frac{\Delta}{1+\sigma_0/\sigma_1\,(s/s_0)^{-\Delta}}
\label{300}
 \eeq
 So, one should expect a growing steepness of the energy dependence for
the total cross section. The value of $r_0$ can be estimated by
demanding that the effective exponent be $\epsilon\approx 0.1$ in
the energy range of fixed target experiments, say at $s\sim
1000\GeV^2$. With $\sigma_0=40\mb$ found in \cite{k3p} one gets
$r_0=0.3\fm$.

Thus, the observed slow rise of the total hadronic cross sections
provides another evidence of the existence of small gluonic spots
with transverse size $r_0\sim 0.3\fm$.

Notice that Eq.~(\ref{300}) may lead to a nonuniversal energy
dependence, which correlates with the magnitude of the total cross
section. Unfortunately, the difference between $\epsilon$ parameters
for $\pi p $ and $pp$ is too small to be observed. Indeed, both
$\sigma_0$ and $\sigma_1$ for $\pi p$ collisions are about a factor
of $2/3$ smaller than for $pp$, and this difference cancels in
(\ref{300}).

For heavier flavors a steeper energy dependence should be expected.
For instance, in the case of $J/\Psi$-proton scattering $\sigma_0$
is so small, that $\epsilon\approx\Delta$. Indeed, data for $J/\Psi$
photoproduction from HERA \cite{psi} show that $\epsilon\approx
0.2$. One should be careful, however, in interpreting the data
within the vector dominance model \cite{hk-vdm}, and also should
remember that Eq.~(\ref{200}) was derived assuming that $r_0$ is
much smaller than the hadronic size, otherwise interferences should
be included.

The $x$-dependence of the proton structure function $F^p_2(x,Q^2)$
at small $Q^2$ also allows to observe the expected nonuniversality
of $\epsilon$.  We expect $\sigma_0$ to fall as
 \beq
\sigma_0(Q^2)\approx \frac{\sigma_0}
{1+Q^2/\Lambda_{QCD}^2}\ ,
\label{400}
 \eeq
 where $\sigma_0\approx \sigma^{\rho p}_{tot} \approx \sigma^{\pi
p}_{tot}$. Correspondingly, $\epsilon$ given by (\ref{300}) first
rises with $Q^2$, then levels off. Of course, at
$Q^2\gg\Lambda^2_{QCD}$, when the transverse size of the $\bar qq$
dipole in the virtual photon $r_T^2\sim 1/(m_q^2+Q^2/4)$ becomes as
small as $r_0^2$, our approach and Eq.~(\ref{300}) break down.

\section{Why \boldmath$\alpha_\Pom^\prime$ is small, but 
\boldmath$\alpha_\Reg^\prime$ is large}\label{slope}

\begin{figure}
 \includegraphics[width=6cm]{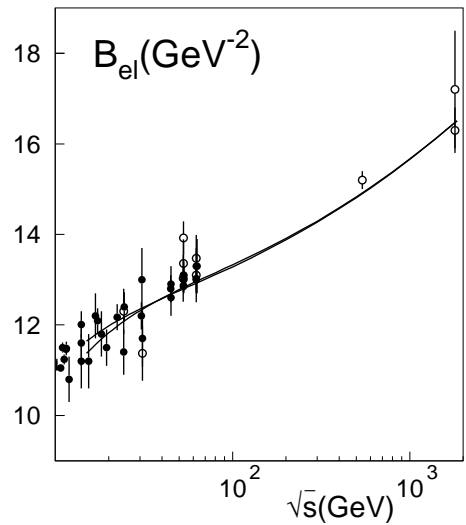}
\caption{\label{slope-fig}
 Slope of elastic $pp$ and $\bar pp$ differential cross sections as
functions of energy. The dashed and solid curve correspond to
Eq.~(\ref{307a}) with $\alpha^\prime_\Pom=0.25\GeV^{-2}$ and to the
slope predicted in \cite{k3p} including the effects of unitarity
saturation, respectively. For references to the depicted
experimental data see \cite{k3p}.}
 \end{figure}

One of the early achievements of the Regge-pole model was the
prediction that the slope of the elastic differential cross section
should rise linearly with $\ln(s)$, as in Eq.~(\ref{307a}). This has
been confirmed by data (see Fig.~\ref{slope-fig})

In terms of the parton model such a shrinkage of the diffractive
cone looks like a result of Gribov's diffusion, i.e. Brownian motion
of cascading partons in impact parameter plane. If gluons are
localized within a small area, i.e. they have a rather large mean
transverse momentum, this may substantially slow down the diffusion.
Indeed, it was found in \cite{k3p} that the result of such diffusion
is rather weak,
 \beq
\alpha_\Pom^\prime =
\frac{\alpha_s}{3\pi}\, r_0^2 \approx 0.1\GeV^{-2}\,.
\label{410}
 \eeq
 This value is quite below the well known one,
$\alpha^\prime_\Pom=0.25\GeV^{-2}$, suggested by data shown in
Fig.~\ref{slope-fig}. However, one should be careful with Regge-pole
approximations in the case of soft hadronic interactions.

\subsection{Effects of saturation in pp elastic scattering}

Since the Pomeron pole amplitude rises with energy as
$f_{el}(b,s)\propto s^\epsilon$, it will eventually hit the
unitarity limit and break the Froissart bound. However, when the
interaction becomes very strong it starts screening not only
specific channels, like in (\ref{306}), but also itself. It is well
known since the 70s how Regge cuts, or absorptive corrections,
restore unitarity \cite{froissaron}, e.g. in the eikonal
approximation one should replace,
 \beq
\Im f_{el}(b,s)\Rightarrow
1-\exp\left[-\Im f_{el}(b,s)\right]\ ,
\label{420}
 \eeq
 Even if the Pomeron slope was zero, $\alpha_\Pom^\prime=0$, when the
amplitude reaches the unitarity bound at small impact parameters,
this area rises as function of energy, i.e. the black disk radius
squared grows proportionally to $\ln(s)$ (amazingly, as for a Regge
pole). The effective slope in the Froissart regime reads,
 \beq
\alpha_{eff}^\prime =
{1\over4}\epsilon B_0=0.22\GeV^{-2}\,,
\label{430}
 \eeq
 which is close to the one observed in data, Fig.~\ref{slope-fig}. We use here
$B_0=8.9\GeV^{-2}$, from Eq.~(\ref{307a}).

\begin{figure}
 \includegraphics[width=8cm]{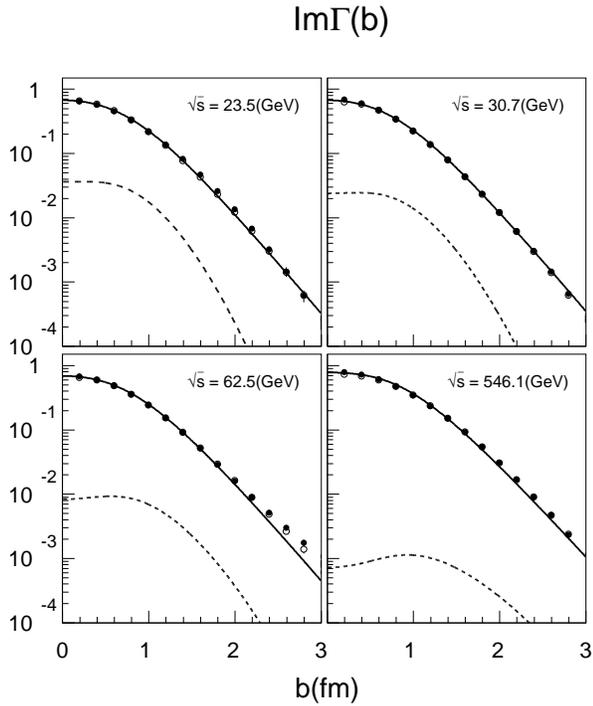}
 \caption{\label{g-b}
 Data points are Fourier transformed data for the differential elastic
cross section \cite{k3p}. Solid curves show the unitarized elastic
amplitude, including Gribov inelastic corrections \cite{gribov}. Dashed curves
correspond to the Reggeon contribution, corrected for absorption.}
 \end{figure}

Proton-proton scattering at the highest accessed energies is close
to this regime, which is something that can be seen from
Fig.~\ref{g-b}, where experimental points are Fourier transformed
data for the elastic differential cross section (see details in
\cite{k3p}).
 The solid curves show the unitarized elastic amplitude, with input the small
Pomeron slope, Eq.~(\ref{410}). The energy dependent slope
corresponding to this amplitude is depicted by the solid curve in
Fig.~\ref{slope-fig}. Apparently, it provides a correct value of
$\alpha^\prime_{eff}$, since agreement with data is good.

\subsection{Far from saturation: electroproduction of vector mesons}

How to disentangle the effects of unitarity saturation and the
genuine Pomeron slope Eq.~(\ref{410})? For this one should switch to
a process which is not affected by the closeness of the unitarity
bound. Elastic scattering of a dipole which is considerably smaller
than the proton would be a proper case. Indeed, neglecting the real
part of the amplitude and assuming exponential $t$-dependence, the
partial amplitude for central collisions ($b=0$) reads,
 \beq
\Im f_{el}(0,s)=\frac{\sigma_{tot}}{4\pi B_{el}}\,.
\label{440}
 \eeq
 For $pp$ collisions this amplitude is slightly below the unitarity
limit, since $\Im f_{el}(b=0)<1$. For $J/\Psi$-proton elastic
scattering, however, the cross section is about an order of
magnitude less than $\sigma^{pp}_{tot}$ \cite{hikt1}, but the slope
is only twice as small as $B^{pp}_{el}$. Thus, we expect $\Im
f^{J/\Psi p}_{el}(0,s)\approx 0.2$ at an energy of the order of
$\sqrt{s}\sim 100\GeV$. This value is safely far below the unitarity
limit, therefore one should expect no influence of saturation on the
value of $\alpha_\Pom^\prime$, which should be as small as given by
Eq.~(\ref{410}).

\begin{figure}
 \includegraphics[width=8cm]{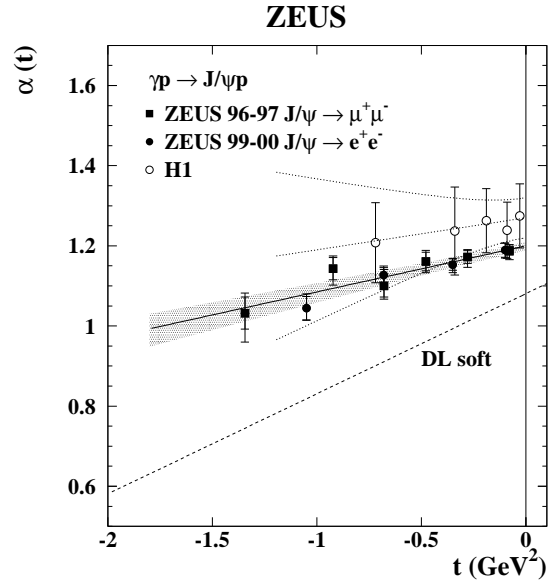}
\caption{\label{psi}
The effective Pomeron trajectory measured in elastic photoproduction
$\gamma p\to J/\Psi p$ \cite{psi}. The slope of the trajectory is
$\alpha^\prime_\Pom=0.115\pm0.018\GeV^{-2}$.}
 \end{figure}

 Therefore to get rid of unitarity corrections one can consider the interaction of a
small dipole with a proton, which can be made small if it involves
heavy flavors, for instance in photoproduction of heavy quarkonium.
In this case the elastic amplitude is also too small to be affected
by unitarity (absorptive) corrections. Then the energy dependence of
the slope may be solely due to the rice of the gluon clouds. Data
for the effective Pomeron trajectory, depicted in Fig.~\ref{psi},
indeed demonstrate a small slope.

\begin{figure}
 \includegraphics[width=7cm]{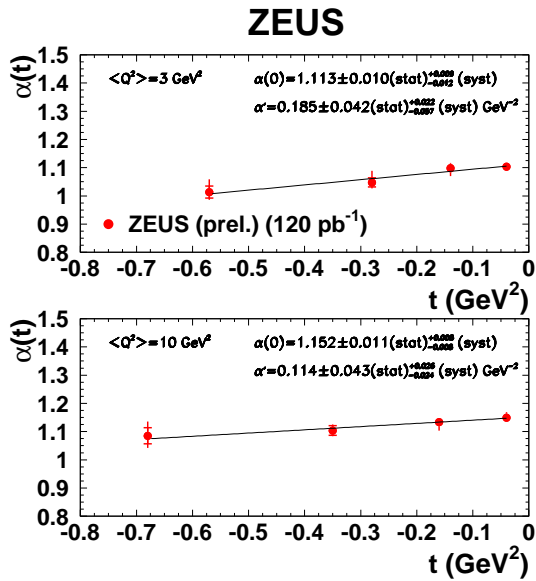}
\caption{\label{rho}
The effective Pomeron trajectory measured in elastic photoproduction
$\gamma p\to \rho p$ \cite{rho-zeus}.}
 \end{figure}

Another way of producing a small dipole is virtual photoproduction of
light vector mesons at high $Q^2$. Indeed, data depicted in Fig.~\ref{rho}
demonstrate a small slope parameter in photoproduction of $\rho$-mesons
at high $Q^2$, which tends to rise towards smaller virtualities.

Here is the list of recent data for the Pomeron slope parameter,
measured in different diffractive channels at HERA.

\begin{itemize}

 \item $\gamma p\to J/\Psi p$, \cite{psi,psi-h1}\,;
 \beqn
\alpha_\Pom^\prime&=&0.115\pm0.018\ ({\rm ZEUS})\,;
\label{450}
\\
\alpha_\Pom^\prime&=&0.164\pm0.028\pm 0.030\ ({\rm H1})\,;
\label{450a}
 \eeqn
 \item $\gamma^* p\to \rho^0 p$, $Q^2=10\GeV^2$ \cite{rho-zeus} (ZEUS)\,;
 \beqn
\alpha_\Pom^\prime&=&0.185\pm0.042^{+0.022}_{-0.057}\ (Q^2=3\GeV^2)\,;\\
\label{460}
\alpha_\Pom^\prime&=&0.114\pm0.043^{+0.026}_{-0.024}\ (Q^2=10\GeV^2)\,;
\label{460a}
 \eeqn
 \item $\gamma^*p\to\phi p$, \cite{phi-zeus} (ZEUS);
 \beq
\alpha_\Pom^\prime=0.08\pm0.09\pm0.08\ (Q^2=5\GeV^2).
\label{470}
 \eeq
 The first and second errors are statistic and systematic
respectively.

\end{itemize}

\subsection{Why \boldmath$\alpha^\prime_\Reg\gg\alpha^\prime_\Pom$?}

The parameter $\alpha^\prime_\Pom$ controlling the energy dependence of the
interaction radius is related to the branching process of gluon raduation by
gluons. Since each impact parameter step of such a branching is small,
$\Delta b\sim r_0$, the resulting expansion is slow and $\alpha^\prime_\Pom$
according to (\ref{410}) is small either. The slope of a Regge trajectory
$\alpha^\prime_\Reg$ is related to gluon emission by a valence quark which
also performs Brownian motion in impact parameters. It is known from data
that $\alpha^\prime_\Reg\approx 1\GeV^{-2}\gg\alpha^\prime_\Pom$.  At first
sight the Brownian motion of a valence quark should be as slow as for gluons,
since if the quark-gluon separation is small for both of them. However, there
are some differences. 

In the case of Pomeron exchange, the gluon should be radiated by a valence
quark with a small fractional momentum $\alpha\ll1$ in order to reach small
$x$. In splitting $q\to qg$ the recoil quark and gluon get impact parameter
shifts $\Delta b_q=\alpha r_T$ and $\Delta b_g=(1-\alpha) r_T$ respectively
where $r_T$ is the final quark-gluon transverse separation. This, the quark
retain its impact parameter, while the gluon makes a transverse step $\Delta 
b_g\approx r_T$.

In the case of Reggeon exchange the valence quark must reach small $x$,
therefore the gluon should be radiated with $\alpha\sim1$, while the
fractional momentum of the quark is very small, $1-\alpha\ll1$. In this case
the quark-qluon separation is shared differently: the gluon shift, $\Delta
b_g=(1-\alpha)r_T$ is vanishing, and the quark makes a maximal step in impact
parameters, $\Delta b_q\approx r_T$. The light-cone $r_T$-distribution
amplitude is also different from Eqs.~(\ref{302}), (\ref{302a}) which assumed
$\alpha\ll1$. For finite $\alpha\to1$ instead of \ref{302a}) one gets,
 \beq
\Psi_{qg}^T(\alpha,\vec r)\Bigr|_{\alpha\to1}={1\over\pi}\,
\sqrt{\frac{\alpha_{s}}{3}}\,  
\chi_f\,\widehat\Gamma\,\chi_i\,K_0(\tau r_T)\ ,
\label{471}
 \eeq
 where $\chi_{i,f}$ are the spinors corresponding to the initial and final 
quarks, and operator $\widehat\Gamma$ has the form \cite{kst1,kst2},
 \beq  
\widehat\Gamma = i\,m_q\,
\vec {e^*}\cdot (\vec n\times\vec\sigma)\,
 + \vec {e^*}\cdot (\vec\sigma\times\vec\nabla)
-i\,\vec {e^*}\cdot \vec\nabla\,.
\label{471a}
 \eeq
 The mean transverse quark-gluon separation here, $\la r_T^2\ra=1/\tau^2$, 
where 
 \beq
\tau^2=\alpha^2m_q^2+(1-\alpha)m_g^2\Bigr|_{\alpha\to1}= m_q^2\,.
\label{471b}
 \eeq
 Thus, differently from Eq.~(\ref{302a}) $\la r_T^2\ra$ depends on the quark, 
rather than gluon mass, i.e. it is quite bigger.
Correspondingly, the slope of the Regge trajectory is larger than the Pomeron 
one, Eq.~(\ref{410}),
 \beq
\frac{\alpha_\Pom^\prime}{\alpha_\Reg^\prime}=
r_0^2\,\Lambda_{QCD}^2\,.
\label{471c}
 \eeq
 Here we assumed that the effective quark mass is of the order of the
inverse confinement radius, $m_q\approx \Lambda_{QCD}$. This, we get the
Reggeon slope $\alpha_\Reg^\prime\approx 1\GeV^{-2}$ in a very good accord
with data. We conclude that the relation between the Reggeon and Pomeron
trajectories slopes is a direct consequence of presence of two scales
in the hadronic structure.

\section{From hard to soft regimes: Two transition scales}\label{hard-soft}

\begin{figure}
 \includegraphics[width=9cm]{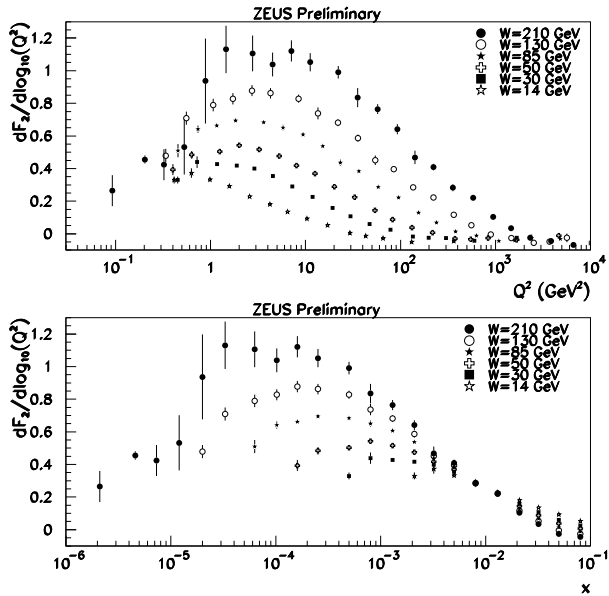}
\caption{\label{caldwell}
Logarithmic scale derivative of the structure function as function of
$Q^2$ and $x$. }
 \end{figure}

The logarithmic $Q^2$-derivative of the structure function
$F_2(x,Q^2)$ at high $Q^2$ serves as a measure of the gluon density
in the proton. Therefore, one should expect a steep growth of this
derivative towards small $x$. Moreover, this growth should slow down
considerably at some very small $x$, due to saturation of the gluon
density, which reaches equilibrium and cannot rise any more. Indeed,
such an expectation was confirmed by ZEUS data \cite{caldwell},
plotted in Fig.~\ref{caldwell} (frequently called Caldwell plot).
The derivative $\partial F_2(x,Q^2)/\partial\ln Q^2$ indeed rises in
accordance with DGLAP evolution down to $x\sim 10^{-3}$, then levels
off an eventually steeply drops at $x < 10^{-4}$. It is not a
surprise that this observation was interpreted right away
\cite{gotsman} as a manifestation of gluon saturation.

One should be cautious, however, treating this as small-$x$ data. Such
very small values of $x$ are reached here not by means of increasing
energy and and keeping $Q^2$ constant, but decreasing $Q^2$.  Certainly,
one can access any tiny value of $x$ by decreasing $Q^2$, but $x$ is not a
proper variable at small $Q^2$ (otherwise one could explore a wonderful
range of small-$x$ physics at Jefferson Lab).

\subsection{Fluctuating photon at the soft transition scale}

In the perturbative regime a virtual photon fluctuates to a $\bar qq$ pair
with a lifetime
 \beq
t_c\sim \frac{2\nu}{M_{\bar qq}^2+Q^2}\,,
\label{472}
 \eeq
 where $\nu$ is the photon energy; $M_{\bar
qq}^2=(k_T^2+m_q^2)/\alpha(1-\alpha)$ is the $\bar qq$ invariant mass
squared; and $k_T$ and $\alpha$ are the quark transverse and fractional
longitudinal momenta respectively.

Perturbatively one treats quarks as free particles. How far can $q$
and $\bar q$ fly apart during the lifetime, Eq.~(\ref{472}) ? The
transverse speed of a quark of energy $\alpha\nu$ is
$v_T^q=k_T/\alpha\nu$ (neglecting the quark mass). Therefore, after
$q$ and $\bar q$ are produced their separation rises during the
lifetime and reaches the value,
 \beq
r_T=t_c(v_T^q+v_T^{\bar q})=
\frac{2k_T}{k_T^2+Q^2\alpha(1-\alpha)}\,.
\label{474}
 \eeq
 The mean value of transverse momentum is given by the energy denominator of the
fluctuation, having the same form as in (\ref{472}), $\la
k_T^2\ra=Q^2\alpha(1-\alpha)$. Then, from Eq.~(\ref{474}) we arrive at the mean $\bar
qq$ separation squared,
 \beq
\la r_T^2\ra=\frac{1}{Q^2\alpha(1-\alpha)}\,.
\label{476}
 \eeq

 Although this heuristic derivation looks classical, it involves quantum
effects via the finite fluctuation lifetime, Eq.~(\ref{472}), which
arises from interferences. A full quantum-mechanical description of
the time evolution of the fluctuation is presented in \cite{kst2}.
The resulting distribution function is well known \cite{bk}, and it
is proportional to the modified Bessel function $K_{0,1}(\epsilon
r)$, where $\epsilon^2=Q^2\alpha(1-\alpha)+m_q^2$. This is simply a
Fourier transform of the energy denominator. The mean transverse
separation of the $\bar qq$ is $\la r_T^2\ra=1/\epsilon^2$, which is
the same as in (\ref{476}) if the quark mass is neglected.

The variation of dipole size with $Q^2$, Eq.~(\ref{476}), generates
the $Q^2$ dependence of the structure function, which describes
quite well the large $Q^2$ data presented in Fig.~\ref{caldwell}.

At high $Q^2$ one can rely on asymptotic freedom and treat quarks as
free particles. This is a key assumption of pQCD. However, in the
limit of real photoabsorption, $Q^2\to0$, this is not appropriate
any more. In this case the quarks are moving within a light-cone
potential controlling the mean transverse separation of $\bar qq$,
which is independent of $Q^2$ (this explains the success of vector
dominance). Thus, the logarithmic $Q^2$ derivative of $F_2$ must
vanish at $Q^2\to 0$, and this indeed is confirmed by the data
depicted in Fig.~\ref{caldwell}.

However, the transition scale for the onset of nonperturbative
effects turns out to be too small, $Q^2\sim
\Lambda^2_{QCD}/\alpha(1-\alpha) <0.16\GeV^2$. This is much smaller
than the values $Q^2\sim 2-4\GeV^2$ where the data depicted in
Fig.~\ref{caldwell} deviate from the pQCD predictions.

\subsection{Semihard transition scale}

The smallness of the mean radius of gluon propagation, $r_0$, should
not affect the proton structure function at small $x\ll1$ and large
$Q^2\gg 4/r_0^2$. In fact, the size of the gluon cloud of a $\bar
qq$ fluctuation of the virtual photon is small anyway, smaller than
$r_0$, since it is cut off by color screening in the same way as in
the case of color transparency. However, when the mean dipole
separation $2/Q$ becomes larger than $r_0$, the interference between
gluon radiation by the $q$ and $\bar q$ weakens, and they acquire
independent (and $Q^2$ independent) gluon distributions. Thus, the
presence of the semihard scale $1/r_0$ leads to a turn-over in the
$Q^2$ dependence of the structure function at $Q^2\sim
4/r_0^2\approx 1.8\GeV^2$.

At smaller $Q^2$, or larger dipole separation, the dipole
cross section is believed to level off. Using the popular parametrization
from \cite{gbw} we get,
 \beqn
\frac{d\sigma(Q^2)}{d\,\ln(Q^2)}&\sim&
\frac{d\sigma(r)}{d\,\ln(r^2)}= {1\over4}\,
\sigma_0\,r^2\,Q_s^2(x)e^{-r^2\,Q_s^2(x)/4}
\nonumber\\ &\sim&
\sigma_0\,\frac{Q_s^2(x)}{Q^2}\,e^{-Q_s^2(x)/Q^2}
\label{478}
 \eeqn
 Thus, the logarithmic derivative stops rising at $Q^2\sim Q^2_s\approx 1\GeV^2$
and turns down. This is in fair agreement with the above estimate.

One can rephrase this in momentum representation: when the $\bar qq$
relative transverse momentum, $k^q_T\sim Q/2$, is much smaller that
the generic transverse momentum of gluons, $k^g_T\sim 1/r_0$, the
variation of $Q$, i.e. of the dipole size, does not affect the gluon
radiation any more. The dipole cross section levels off at large
separation, leading to a reduction of the logarithmic derivative
Eq.~(\ref{478}).

The observed behavior of the logarithmic $Q^2$ derivative of
$F_2(x,Q^2)$ has been interpreted in terms of low-$Q^2$ behavior,
rather than low-$x$, in Ref. \cite{kaidalov-1}. However, a semi-hard
scale $Q_0^2=2\GeV^2$ was in this case introduced, with no
motivation.

\subsection{Variation of the energy dependence with the scale}

\begin{figure}
 \includegraphics[width=7cm]{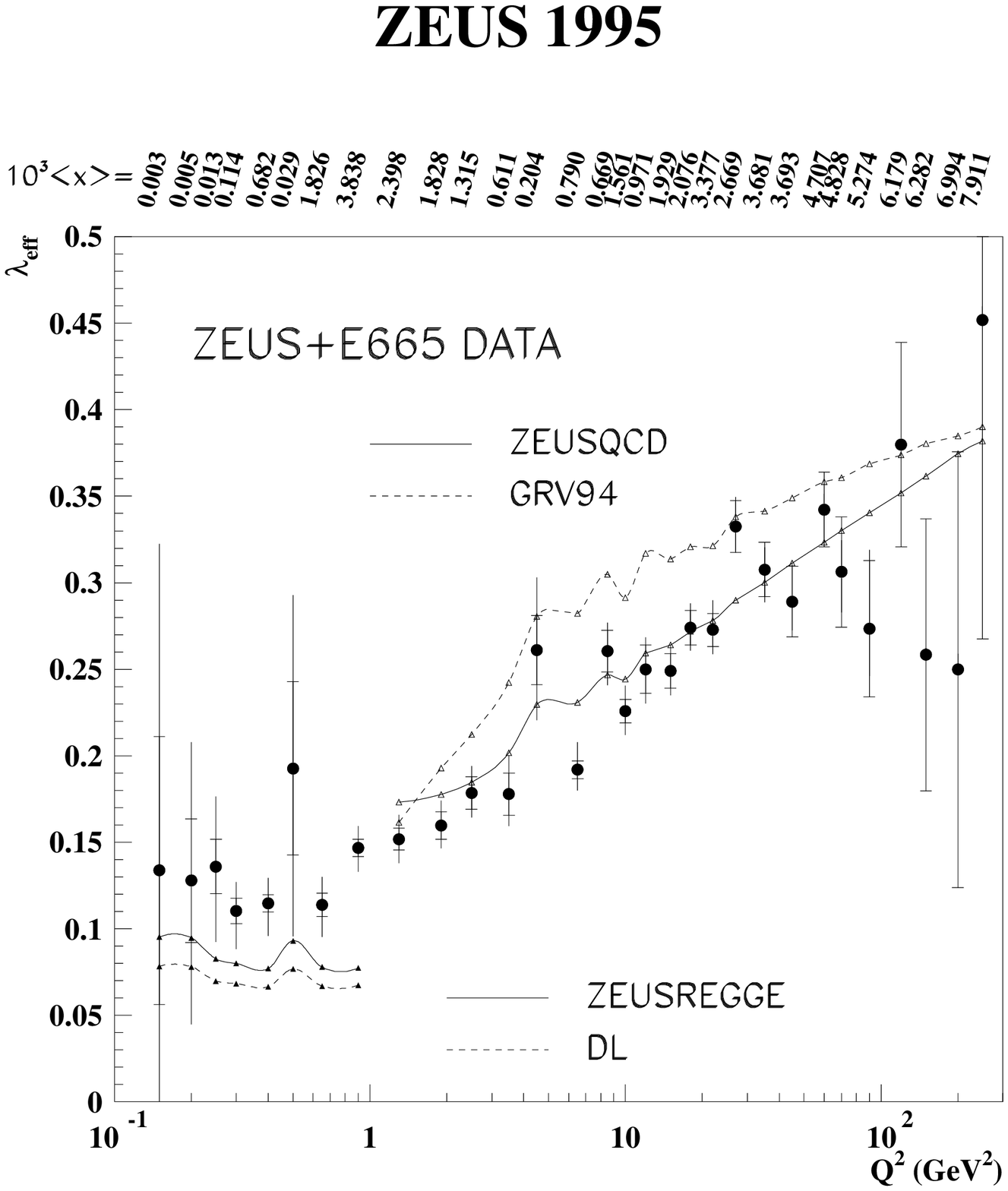}
\caption{\label{lambda}
 $\lambda_{eff}=d\ln F_2/d\ln(1/x)$ as a function of $Q^2$, calculated in
\cite{lambda} by fitting $F_2=A x^{-\lambda_{eff}}$ to ZEUS and E665
data with $x<0.01$. The DL and GRV94 calculations are from Ref.
\cite{dltwo} and the GRV94 NLO fit \cite{grv94}, respectively. }
 \end{figure}

One of the main results of HERA is the observation of a steep rise
with $Q^2$ of the energy slope
$\lambda_{eff}(Q^2)=d\ln(F_2)/d\ln(s)$, which is well explained by
DGLAP evolution. On the other hand, one expects $\lambda$ to
approach the known hadronic value $\lambda=0.1$ in the soft limit
$Q^2\to0$. At which value of $Q^2$ does this transition happens ?
The change of regime should occur for the same reason as it was
discussed above for the logarithmic $Q^2$ derivative, namely, when
the mean $\bar qq$ separation exceeds the mean sized of the gluonic
clouds, $r_0\approx 0.3\fm$. Therefore it is expected to happen at
the same semi-hard scale. ZEUS data for $\lambda_{eff}(Q^2)$
presented in Fig.~\ref{lambda} confirm this.

\section{Fermi motion of quark and gluons}
\label{primordial}

At a soft scale one does not resolve the gluonic structure of a hadron,
but only the valence quarks. The mean transverse Fermi momentum of these
quarks is small,
 \beq
k^v_T\sim {1\over r_h}\sim\Lambda_{QCD}\,.
\label{480}
 \eeq

At higher scale relevant to hard reactions one can resolve the
structure of the valence quarks, i.e. the presence of gluons and sea
quarks. Since those are located at small separations, $\sim r_0$,
from the valence quark, both have more intensive intrinsic Fermi
motion,
 \beq
\la k_0^2\ra\sim{1\over r_0^2}\,.
\label{490}
 \eeq
 Such a large primordial momentum should affect the onset of the hard
regime in the particular process.

\subsection{Cronin effect}

A projectile parton propagating through a nucleus experiences
multiple interactions increasing its transverse momentum. Then the
parton participating in a hard collision inside the nucleus has an
increased transverse momentum compared to Eq.~(\ref{490}), which
corresponds to the interaction with a free proton,
 \beq
\la k_A^2\ra=\la k_0^2\ra +
\Delta k^2\,,
\label{500}
 \eeq
 where $\Delta k^2$ is the nuclear broadening.

This observation helps to understand the Cronin effect \cite{cronin}
of nuclear enhancement of hadron production with high $p_T$.
Apparently, the strength of the effect depends on the relative
values of the two terms in (\ref{500}). In the limit of a weak
primordial motion the effect should be strongest, while in the case
of $\la k_0^2\ra\gg \Delta k^2$ the effect will disappear.

\begin{figure}[htb]
\begin{center}
 \includegraphics[width=7cm]{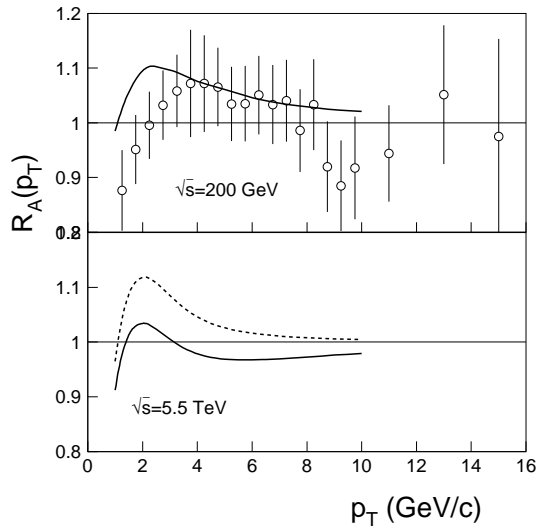}
 \caption{ \label{fig-cronin} Nucleus-to-proton cross section ratio for
pion production, versus $p_T$. Dashed and solid curves correspond to
calculations without or with gluon shadowing \cite{knst}}
 \end{center}
 \end{figure}

A rather strong Cronin effect was observed in fixed target experiments,
where production of high-$p_T$ hadrons is dominated by scattering of
valence quarks \cite{knst} which have a small primordial $k_T\sim
1\fm^{-1}$. One can access the gluons only at sufficiently high energies.
Relying on the above consideration, a very weak Cronin enhancement was
predicted in \cite{knst} at $\sqrt{s}=200\GeV$, as is depicted in
Fig.~\ref{fig-cronin}. A several times stronger effect was predicted in
\cite{wang}\footnote{The extremely strong gluon shadowing implemented into
the HIJING model is ruled our by the recent LO \cite{ehks} and NLO
analyses \cite{florian} of DIS data.}, and a suppression, rather than
enhancement, was the expectation of the color glass condensate (CGC) model
\cite{klm}. The latest data from the PHENIX experiment at RHIC, depicted
in Fig.~\ref{fig-cronin}, confirms the prediction of \cite{knst}.

It is worth mentioning that interferences, which are usually assumed to
be negligible for hard reactions on nuclei, become important if the
collision energy is sufficiently high. In this case the process is
called coherent. The mechanisms of the Cronin enhancement are quite
different in these two limiting cases. In the case of incoherent
hard interaction the incoming projectile partons (mainly valence
quarks) experience initial state multiple soft rescattering leading
to a high-$p_T$ enhancement, and the projectile gluons located in
small spots are not resolved. If, however, the energy is
sufficiently high, the lifetime of the hard fluctuation exceeds the
nucleus size and the process becomes coherent. In this case gluons
are well resolved, since they dominate the production of high-$p_T$
partons. Such a coherent regime is relevant for hadron production at
medium large $p_T$ at RHIC, and it dominates a large range of
$p_T$ at LHC energies. Besides, in the latter case the Cronin effect
is substantially reduced by shadowing (see next Sect.), as is
demonstrated in Fig.~\ref{fig-cronin}.

\begin{figure}[htb]
\begin{center}
 \includegraphics[width=7cm]{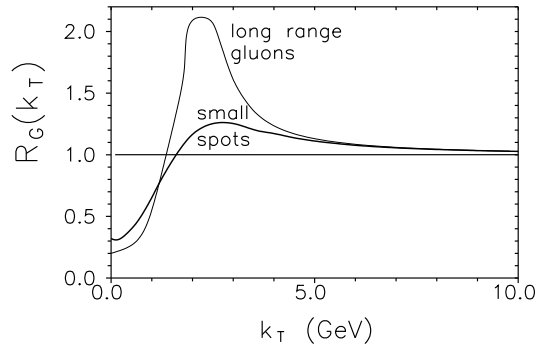}
 \caption{ \label{ratio}
Nucleus-to-proton cross section ratio for gluon radiation as function of 
$p_T$. The upper and bottom curves correspond to mean gluon propagation distances 
$1\fm$ (long range) and $0.3\fm$ (small spots) respectively
\cite{kst1}. }
 \end{center}
 \end{figure}

In the coherent regime the interaction resolves the gluons and their primordial 
transverse momentum is of great importance: the larger it is, the smaller is the 
Cronin enhancement. This is demonstrated on Fig.~\ref{ratio} for coherent gluon
radiation by a quark propagating through a nucleus. 
 The depicted ratio of the radiation cross sections on nucleus-to-proton targets
is calculated with different assumptions about the mean propagation length of 
gluons. The upper curve corresponds to a long range propagation $\sim 1\fm$, while 
the bottom one assume that gluons are confined within small spots $\sim 0.3\fm$.
In the latter case the gluon primordial momentum is larger, correspondingly the 
Cronin enhancement is much weaker in a good accord with RHIC data 
shown in Fig.~\ref{fig-cronin}.

\subsection{Other processes}

The mean primordial transverse momentum of projectile partons can be
generated perturbatively. Namely, higher order processes include
gluon bremsstrahlung, which leads to increasing transverse momenta
of partons radiating gluons. Nevertheless, NLO calculations fail to
explain data on Drell-Yan reactions \cite{dy}, direct photon
\cite{apanasevich} and heavy flavor production \cite{hf}, unless an
additional primordial transverse momentum $k_T$ is introduced. All
these reactions demand a mean primordial momentum, $\la
k_T\ra\approx 1\GeV$.

\begin{figure}[htb]
\begin{center}
 \includegraphics[width=7cm]{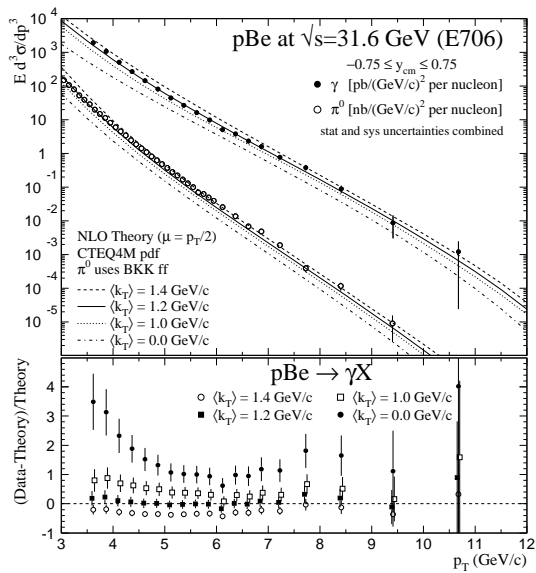}
 \caption{ \label{kt-photon}
Top: the photon and $\pi^0$ production cross sections from the E706
experiment at $\sqrt{s}=31.6\GeV$, compared to $k_T$-corrected NLO
calculations \cite{apanasevich}. Bottom: the ratio
(Data-Theory)/Theory for direct photon production. Theory is the NLO
calculations with primordial parton momentum $\la k_T\ra$.}
 \end{center}
 \end{figure}

Comparison of NLO calculations and data \cite{apanasevich}, for both
direct photons and neutral pion production, is depicted in
Fig.~\ref{kt-photon}.
 The primordial momentum $\la k_T\ra\approx 1.2 GeV$ seems to provide the best
description of data. This value is somewhat larger than the semihard
scale under discussion, but it includes contributions from higher
than NLO terms. Besides, in this calculations there is no strict
border between the soft part of gluon radiation responsible for $\la
k_T\ra$, and the hard part ascribed to the perturbative NLO part.

Notice that in the dipole description of direct photon
\cite{kst1,amir} and Drell-Yan pair \cite{joerg} production there is
no need (no uncertainty either) to introduce any primordial
momentum, beyond a soft one related to the hadron size. The semihard
primordial momentum is included by default in the phenomenological
dipole cross section.

A similar value of the primordial momentum was extracted from
azimuthal angle correlations measured in high-$p_T$ di-hadron
production at RHIC \cite{rak},
 \beq
\sqrt{\la k_T^2\ra}= 1.28\pm0.06\GeV\,.
\label{510}
 \eeq
 However, in this case no NLO calculation has been performed, and it is not clear
which part of this primordial momentum has a nonperturbative origin.

Another observable sensitive to the primordial parton motion is the
so called seagull effect, the dependence of mean transverse momentum
of produced hadrons on Feynman $x_F$. The observed shape of the
$x_F$-dependence is a result of fragmentation of forward jets
\cite{close}. Their transverse momenta include the primordial one.
In was found \cite{seagull} that data demand a rather large $\la
k_T\ra \sim 0.5-1\GeV.$

\section{Nuclear glue: saturated or dilute?}\label{nuclei}

\begin{figure}
 \includegraphics[width=5cm]{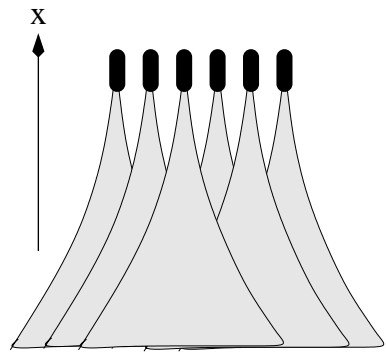}
\caption{\label{clouds}
Logarithmic scale derivative of the structure function as function of
$Q^2$ and $x$. }
 \end{figure}

Gluon clouds originated from different nucleons can overlap in
longitudinal direction at small $x$, since the internucleon spacing
contracts $\propto 1/E$, while gluon clouds at small $x$ shrink as
$1/xE$. This is illustrated in Fig.~\ref{clouds}.
 Since gluons overlap they can interact and fuse, and this leads to a
reduction of the number of gluons in nuclei at small $x$. This phenomenon
is called gluon shadowing.

\begin{figure}
 \includegraphics[width=7cm]{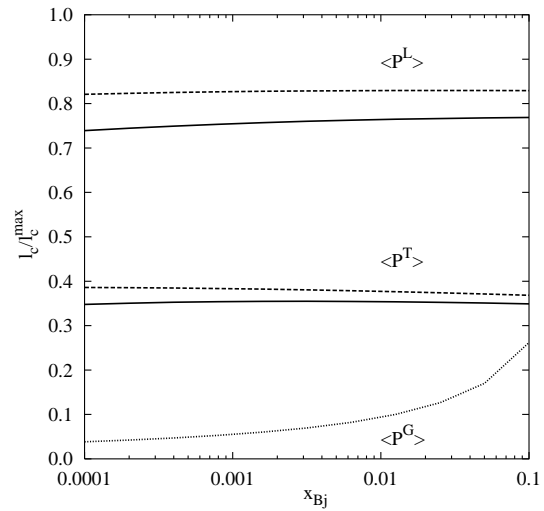}
\caption{\label{lc} Bjorken $x$ dependence of the factor $P$ in
Eq.~(\ref{600}) for quarks and longitudinal (two upper curves) and
transverse (next couple of curves) photons. Solid and dashed curves
are calculated \cite{krt2} for $Q^2=4$ and $40\GeV^2$ respectively.
The bottom curve represents $P$ for gluons.}
 \end{figure}

It turns out that the smallness of the gluonic spots affects the
longitudinal overlap. First it makes the clouds effectively shorter
in the longitudinal direction. Indeed, the coherence length which
controls the overlap of the clouds is shorter, since gluons with an
enhanced transverse motion are effectively heavier. The coherence
length has the form,
 \beq
l_c=\frac{P}{xm_N}\equiv Pl_c^{max}\,,
\label{600}
 \eeq
 where the factor $P$ is usually assumed to be $p\approx 1/2$. However,
even for quarks $P$ depends considerably on the polarization of the
virtual photon, but for gluons it is an order of magnitude shorter
\cite{krt2}. One can see this in Fig.~\ref{lc}.
 Thus, gluons need much smaller $x$ than quarks to overlap in longitudinal
direction, i.e. the onset of gluon shadowing occurs at quite smaller $x$
than for quark shadowing.

 However, it is not sufficient for gluons to overlap in longitudinal
direction, in order to interact they also have to overlap in
transverse plane. If gluon clouds are as big as the proton, the mean
number of other clouds seen by a particular one is $\pi R_p^2 \la
T_A\ra$, where the mean nuclear thickness is $\la T_A\ra=1/A\int
d^2b\,T_A^2(b)$, and
$T_A(b)=\int_{-\infty}^{\infty}dz\,\rho_A(b,z)$. For lead $\la
T_A\ra=1.35\fm^{-2}$, so the mean number of such overlaps is rather
large, about 4.

\begin{figure}[h]
 \includegraphics[width=7cm]{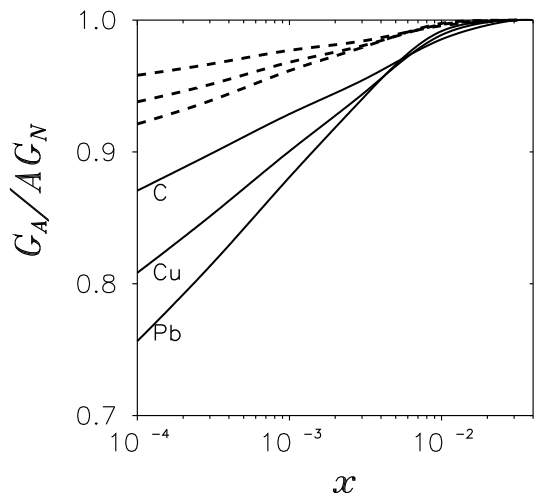}
\caption{\label{glue-shad} Ratio of the gluon distribution functions
in nuclei (carbon, copper and lead) and nucleons versus Bjorken $x$
at $Q^2 = 4\,GeV^2$ (solid curves) and $40\,GeV^2$ (dashed curves).}
 \end{figure}

On the other hand, if the quark-gluon separation is as small as $r_0$,
the mean number of other dipoles overlapping transversely with this one is,
 \beq
\la n\ra=\frac{3\pi r_0^2}{4}\,\la T_A\ra
\approx 0.3\,.
\label{610}
 \eeq
 This shows that even at small $x$ the overlap of gluons is very small,
therefore shadowing is weak. An example is depicted in
Fig.~\ref{glue-shad}

Experimental information about gluon shadowing is extremely scarce.
Only one experiment, NMC \cite{nmc}, was able to detect a variation
with $Q^2$ of nuclear effects in DIS at small $x$. Leading order
DGLAP analysis \cite{eks,hkm} of this data were unable to extract
gluon shadowing. Only NLO analysis by De Florian and Sassot
\cite{florian} turned out to be sensitive to gluon shadowing, which
was found to be very weak, as predicted in \cite{kst2}, and even
somewhat weaker.

Unfortunately, this evidence of gluon shadowing is based on the
results of only one experiment. New measurements are desperately
needed, and the eRHIC collider would offer a wonderful opportunity
for this physics.

\section{Summary}

Presence of the semi-hard scale in hadrons, corresponding to
distances as small as $0.3\fm$ leads to numerous observable effects. 
It helps to explain from a common point of view a series of puzzling features
observed in particle collisions at high-energies.
\begin{itemize}
 \item
 Smallness of the triple-Pomeron coupling, which means that
diffractive gluon radiation is suppressed.
 \item
 $t$-independence of the triple-Pomeron coupling.
 \item
 Small value of the Pomeron intercept, $\alpha_{\Pom}(0)-1=0.1$.
 \item
 Small value of of the Pomeron trajectory slope,
$\alpha^\prime_\Pom\approx 0.1\GeV^{-2}$ observed in exclusive
electroproduction of vector mesons. At the same time the unitarity
saturation effects lead to a larger value,
$\alpha^\prime_{eff}=0.25\GeV^{-2}$ in $pp$ collisions. Although Brownian
motion of gluons in impact parameters is very slow, valence quarks
emitting gluons move much faster, and Eq.~(\ref{471c}) well reproduces the
large slope of Regge trajectories $\alpha^\prime_\Reg\approx1\GeV^{-2}$.
This relation, Eq.~(\ref{471c}), is a direct consequence of the two-scale 
structure of hadrons.
 \item
 The two hadronic scales can be seen in the transition from hard to soft
regimes in deep-inelastic scattering.  The logarithmic derivative $\partial
F_2/\partial\ln(Q^2)$ vanishes at very small $Q^2<4\Lambda_{QCD}^2$, but
deviates from the expected DGLAP behavior at much higher values $Q^2\sim
2\GeV^2$.
 \item
 Presence of the semihard scale in the primordial transverse momenta of
partons leads to a very weak nuclear (Cronin) enhancement of high-$p_T$
hadrons observed in data at RHIC. This means that color glass condensate is a
rather weak effect in the available energy range.
 \item
 Other hard reactions sensitive to the primordial parton motion (direct
photon, Drell-Yan dileptons, heavy flavors, back-to-back di-hadrons, seagull
effect, etc.) also demand a large transverse momenta of the projectile
partons, which cannot be explained by NLO calculations.
 \item
 NLO analysis of nuclear structure functions demonstrates a very weak gluon
shadowing which is a clear result of a strong gluon localization. This result
bridges to the observed weak diffractive excitation of gluonic degrees of
freedom. Indeed, nuclear shadowing is directly related to diffraction via
Gribov's inelastic shadowing \cite{gribov}.

\end{itemize}

Notice that for most experimental facts presented in this list no alternative
explanation has been proposed so far. Therefore they maybe considered as
solid experimental evidences in favor of presence of a semi-hard scale in
hadronic structure. Still, this important issue demands further study, and
new precise data are desperately needed. In this respect a planned
electron-nucleus collider (eRHIC) will be a wonderful facility for these
researches.

\bigskip

{\bf Acknowledgments:} We are thankful to Hans-J\"urgen Pirner for interesting
discussions. This work was supported in part by Fondecyt (Chile) grants numbers
1050589 and 1050519, and by DFG (Germany) grant PI182/3-1.


\begin{thebibliography}{99}

\bibitem{bfkl} L.N.~Lipatov, Sov.\ J.\ Nucl.\ Phys.\ {\bf 23}, 338 (1976);
V.S.~Fadin, E.A.~Kuraev and L.N.~Lipatov, Phys.\ Lett.\ B {\bf 60}, 50 (1975);
I.I.~Balitsky and L.N.~Lipatov, Sov.\ J.\ Nucl.\ Phys.\ {\bf 28}, 882 (1978)  ;
\ JETP Lett.\ {\bf 30}, 355 (1979).

\bibitem{kst2} B. Z. Kopeliovich, A. Sch\"afer, and A. V. Tarasov, Phys. Rev. D
\textbf{62}, 054022 (2000).

\bibitem{lattice} A.~DiGiacomo and H.~Panagopoulos, Phys.  Lett. B {\bf 285}
(1992) 133.

\bibitem{shuryak} T. Sch\"afer, E.V. Shuryak, Rev. Mod. Phys. {\bf 70}, 323
(1998).

\bibitem{shuryak-zahed} E. Shuryak and I.~Zahed, Phys. Rev. D69 (2004) 014011.

\bibitem{braun} V.M.~Braun, P.~G\'ornicki, L.~Mankiewicz, Phys. Lett.
B302 (1993) 291.

\bibitem{knst} B.Z. Kopeliovich, J.~Nemchik, A.~Sch\"afer, and A.V.~Tarasov,
Phys. Rev. Lett. {\bf 88}, 232303 (2002).


\bibitem{kp} B.Z. Kopeliovich and B.~Povh, J. Phys. G30 (2004) S999.

\bibitem{kps} B.~Z.~Kopeliovich, B.~Povh and I.~Schmidt,
Nucl.\ Phys.\  A {\bf 782}, 24 (2007).

\bibitem{kklp} Yu.M.~Kazarinov, B.Z. Kopeliovich, L.I.~Lapidus, and
I.K.~Potashnikova, Sov.  Phys. JETP {\bf 70} (1976) 1152.

\bibitem{kaidalov} A.~B.~Kaidalov, Phys.\ Rept.\ {\bf 50}, 157 (1979).

\bibitem{schlein1} A.~Brandt {\it et al.}  [UA8 Collaboration],
Eur.\ Phys.\ J.\  C {\bf 25}, 361 (2002).

\bibitem{zkl} B.Z.~Kopeliovich, L.I.~Lapidus, and A.B.~Zamolodchikov, Sov.
Phys. JETP Lett. {\bf 33}, 595 (1981); Pisma v Zh. Exper. Teor.  Fiz. {\bf
33}, 612 (1981).

\bibitem{ingelman} G.~Ingelman and K.~Prytz, Z. Phys. C {\bf 58}, 285
(1993).

\bibitem{kst1} B.Z.~Kopeliovich, A.~Sch\"afer and A.V.~Tarasov,
Phys. Rev. {\bf C59},1609 (1999).

\bibitem{qm04} B.~Z.~Kopeliovich and B.~Povh, J.\ Phys.\ G {\bf 30}, S999
(2004).

\bibitem{amaldi} U.~Amaldi and K.R.~Schubert, Nucl. Phys. {\bf B166}, 301
(1980).

\bibitem{k2p} B. Z. Kopeliovich, B.~Povh and E.~Predazzi, Phys. Lett. {\bf
B405}, 361 (1997).

\bibitem{k3p} B.Z. Kopeliovich, I.K.~Potashnikova, B.~Povh, and
E.~Predazzi, Phys. Rev. Lett. 85 (2000) 507; Phys. Rev. D63 (2001) 054001.

\bibitem{glm} E.~Gotsman, E.M.~Levin, and U.~Maor, Z. Phys. C{\bf 57}, 677
(1993); Phys. Rev. D{\bf 49}, 4321 (1994); Phys. Lett. B{\bf 353}, 526
(1995); Phys.  Lett. B{\bf 347}, 424 (1995).

\bibitem{dino} K.Goulianos, J.~Montanha, Phys. Rev. {\bf D59}, 114017
(1999)

\bibitem{peter} S.~Erhan and P.E.~Schlein, Phys. Lett. {\bf B427} (1998)
389.

\bibitem{pdg} Review of Particle Physics,
S. Eidelman et al., Phys. Lett. B592, 1 (2004).

\bibitem{psi} S.~Chekanov {\it et al.} [ZEUS Collaboration],
Nucl.\ Phys.\ B {\bf 695}, 3 (2004).

\bibitem{hk-vdm} J.~H\"ufner and B.~Z.~Kopeliovich, Phys.\ Lett.\ B {\bf 426},
154
(1998).

\bibitem{froissaron} M.~S.~Dubovikov, B.~Z.~Kopeliovich, L.~I.~Lapidus and
K.~A.~Ter-Martirosian, Nucl.\ Phys.\ B {\bf 123}, 147 (1977).

\bibitem{gribov} V.N.~Gribov, Sov. Phys. JETP {\bf 56}, 892 (1968).

\bibitem{hikt1} J.~H\"ufner, Yu.~P.~Ivanov, B.~Z.~Kopeliovich and A.~V.~Tarasov,
Phys.\ Rev.\  D {\bf 62}, 094022 (2000).

\bibitem{psi-h1} A.~Aktas {\it et al.} [H1 Collaboration], Eur.\ Phys.\ J.\ C
{\bf 46}, 585 (2006).

\bibitem{rho-zeus} S.~Chekanov et al.  [ZEUS Collaboration],
  DESY-07-118, arXiv:0708.1478 [hep-ex].

\bibitem{phi-zeus} S.~Chekanov {\it et al.} [ZEUS Collaboration], 
electroproduction of Phi mesons at HERA,'' Nucl.\ Phys.\ B {\bf 718}, 3 (2005).

\bibitem{gotsman} E.~Gotsman, arXiv:hep-ph/9906436.

\bibitem{bk} J.M. Bjorken, J.B. Kogut, and D.E. Soper, Phys. Rev. D {\bf 3},
1382 (1971).

\bibitem{gbw} K.~Golec-Biernat and M.~W\"usthoff, Phys. Rev. D{\bf 59}, 014017
(1999).

\bibitem{kaidalov-1} C.~Merino, A.~B.~Kaidalov and D.~Pertermann,
arXiv:hep-ph/9911331.

\bibitem{caldwell} A. Caldwell, DESY Theory Workshop, DESY, Hamburg (Germany),
October 1997; J. Breitweg et al (ZEUS Collaboration), Eur. Phys. J. C{\bf 7},
609 (1999); B.~Foster, Eur.\ Phys.\ J.\ direct C {\bf 4S1}, 37 (2002).

\bibitem{lambda} J.~Breitweg {\it et al.} [ZEUS Collaboration], Eur.\ Phys.\
J.\ C {\bf 7}, 609 (1999)

\bibitem{dltwo} A.~Donnachie \& P.~Landshoff, Z. Phys. C{\bf 61}, 139 (1994).

\bibitem{grv94} M.~Gl\"{u}ck, E.~Reya \& A.~Vogt, Z. Phys. C{\bf 67}, 433
(1995).

\bibitem{cronin} D.~Antreasyan et al., Phys. Rev. {\bf D19}, 764 (1979)

\bibitem{wang} X.~N.~Wang, Phys.\ Rev.\ C {\bf 61}, 064910 (2000).

\bibitem{ehks} K.~J.~Eskola, H.~Honkanen, V.~J.~Kolhinen and C.~A.~Salgado,
Phys.\ Lett.\  B {\bf 532}, 222 (2002).

\bibitem{florian} D.~de~Florian and R.~Sassot, Phys. Rev. D{\bf 69}, 074028
(2004).

\bibitem{klm} D. Kharzeev, E. Levin, L. McLerran, Phys. Lett. B{\bf 561}, 93
(2003).

\bibitem{dy} D. C. Hom, et al., Phys. Rev. Lett. 37, 1374 (1976); D.
M. Kaplan, et al., Phys. Rev. Lett. {\bf 40}, 435 (1978).

\bibitem{apanasevich} L. Apanasevich, et al., Phys. Rev. Lett. 81, 2642 (1998);
L. Apanasevich, et al., Phys. Rev. D {\bf 59}, 074007 (1999).

\bibitem{hf} M.~L.~Mangano, P.~Nason and G.~Ridolfi,
Nucl.\ Phys.\  B {\bf 373}, 295 (1992).

\bibitem{amir} B.~Z.~Kopeliovich, A.~H.~Rezaeian, H.~J.~Pirner and I.~Schmidt,
arXiv:0704.0642 [hep-ph], to appear in Phys. Lett. B.

\bibitem{joerg} B. Z. Kopeliovich, J. Raufeisen and A. V. Tarasov, Phys.
Lett. B{\bf 503}, 91 (2001).

\bibitem{rak} J.~Rak, Talk at International Conference on High-Energy
Interactions, Herlany, Slovakia, 22-27 Sep. 2002 (arXiv:nucl-ex/0306031).

\bibitem{close} F.~E.~Close, F.~Halzen and D.~M.~Scott, Phys.\ Lett.\ B {\bf
68}, 447 (1977).

\bibitem{seagull} NA22 Collaboration, N.M.~Agababyan et al., Phys. Lett.  B{\bf
320}, 411 (1994).

\bibitem{krt2} B.Z.~Kopeliovich, J.~Raufeisen and A.V.~Tarasov, Phys. Rev. C{\bf
62}, 035204 (2000).


\bibitem{nmc} P. Amaudruz et. al., Nucl. Phys. B441, 3 (1995); M. Arneodo et
al., B{\bf 441}, 12 (1995).

\bibitem{eks} K.J. Eskola, V.J. Kolhinen, P.V. Ruuskanen, Nucl. Phys. B535, 351
(1998); K. J. Eskola, V. J. Kolhinen, C. A. Salgado, Eur. Phys. J. C{\bf 9}, 61
(1999).

\bibitem{hkm} M. Hirai, S. Kumano, M. Miyama, Phys. Rev. D 64 034003 (2001).

\bibitem{5} F.~Hautmann, D.E.~Soper, Phys. Rev. D63 (2000) 011501.

\bibitem{mv} L.~McLerran and R.~Venugopalan, Phys. Rev. D49
(1994) 2233.


\end{thebibliography}
\end{document}